\documentclass[12pt,preprint]{aastex}

\shorttitle{Five-cluster photometry}
\shortauthors{Taylor \& Joner}

\begin{document}

\title{SELECTING HIGH-PRECISION PHOTOMETRY ON UNIFORM ZERO POINTS FOR FIVE BENCHMARK GALACTIC 
CLUSTERS}

\author{Benjamin J.\ Taylor and Michael D.\ Joner}
\affil{Department of Physics and Astronomy, N283 ESC, Brigham Young University,
   Provo, UT 84602-4360}

\begin{abstract}
This paper reviews results from two projects designed to yield photometry on uniform zero points for 
five clusters--Coma, the Hyades, M67, NGC 752, and Praesepe.  Contributing papers for a project on 
Cousins {\it VRI\/} photometry and a project on Str{\" o}mgren-$\beta$ photometry are listed.  Results 
of zero point tests of the photometry are reviewed, and their character is found to be satisfactory at 
the level of a few mmag.  Responses to extant criticisms of the photometry are offered, and a section 
on $B-V$ photometry for the five clusters is included.  Because the results of the projects suggest 
that certain changes should be made in current perspectives on photometry, those changes are reviewed.  
Finally, suggestions are made about future uses of data from the projects.
\end{abstract}


\section{INTRODUCTION}

In 1985 and 1992, we started ``zero point'' projects designed to yield photometry on uniform zero 
points for a number of galactic clusters.  For five of them--Coma, the Hyades, M67, NGC 752, and 
Praesepe--complete results have now been published.  Judging from ADS searches, however, at least some 
of those results are not as well known as they might be.  In addition, readers of the papers from 
the projects would probably find it hard to put them in perspective.

Unavoidably, the papers are very detailed.  As a result, potential users of the photometry from the 
projects would not find it easy to read straight through them.  Even with the help of their abstracts, it 
might be difficult for users to decide just how convincing each paper is.  Currently, moreover, only fairly 
detailed comparisons among the papers could reveal which of them contain up-to-date data, which contain only 
superseded data, and which include discussions that are still topical.  At present, it appears that at least 
one important project paper has been overlooked altogether.  Finally, there is currently no published 
assessment of the coherence (or lack of coherence) of published photometry that has been surveyed during 
the course of the projects.

This paper addresses all of these problems.  All papers produced by the two projects are listed, and potential 
users of the data are directed to papers that contain current results.  Outcomes of zero point tests are 
summarized, and sources of more detailed information are cited.  Because the conceptual foundation of 
the projects is controversial, existing criticisms of the photometry and our responses are discussed at some 
length.  Given the results of the projects, some changes in outlook by photometric astronomers seem to be called 
for, and these are discussed.  Finally, suggestions are made about possible use of project data.

Readers are invited to begin by reading \S 2 and a literature source cited there.  Subsequent sections are 
concerned with Cousins {\it VRI\/} colors (\S 3), associated $V$ magnitudes (\S 4), $B-V$ colors (\S 5),
Str{\" o}mgren-$\beta$ colors (\S 6), and a set of perspectives (\S 7).  Readers who consult \S 3 are invited 
to look at \S\S 3.1, 3.2, 3.8, 3.9, and the subsections concerning clusters of interest to them.

\section{THE TWO PROJECTS: BASICS}

\subsection{Quantities measured, instrumental systems, and published papers}

For one of the projects, the focus is on Str{\" o}mgren-$\beta$ measurements.  The other project is concerned 
primarily with values of Cousins {\it VRI}.  Papers from the latter project can be divided into a ``Landolt'' 
group and an ``SAAO'' group.  Measurements in the Landolt group have been reduced to the \citet{l83} system of 
standard stars.  For this group, the principal source of current data is a paper co-authored with Elizabeth J.\ 
Jeffery \citep{tjj08}.   

Measurements in the ``SAAO'' group have been obtained at the South African Astronomical Observatory.  Here, our 
collaborators have been C.\ David Laney and Francois van Wyk.  To reduce measurements in this group, standard stars 
on the native Cousins system that have been calibrated at SAAO have been used.  In addition to Cousins photometry, 
the papers in this group report values of $B-V$.

The character of the SAAO instrumental system plays a key role in our work (see \S 3).  That system has a history 
of yielding stable transformations to the Cousins system (see, for example, \S 2 of \citealt{jtlw06}).  Moreover, the 
SAAO system has been used extensively to establish Cousins {\it VRI\/} standards in the southern hemisphere (see, for 
example, \citealt{mc89,kw98,kk02}).  For these reasons (see also \S 3.2), the SAAO system is regarded here as the 
most authoritative system of its kind.  

All but one of the papers produced by the projects are listed in Table 1.  A paper by \citet{tj96} that 
compares the Cousins and \citet{l83} {\it VRI\/} systems is not listed in the table because it contains no cluster
photometry.  However, that paper should be regarded as part of the {\it VRI\/} project (see \S 3.1).  We also note 
that the first SAAO paper listed in Table 1 has been cited several times (see, for example, \citealt{tate09} and 
\citealt{clhw11}).  However, TJJ has not been cited often, and it appears that its existence is not widely known.  
Potential users of project {\it VRI\/} data are therefore invited to adopt TJJ data for the Hyades, M67, Coma, and/or 
NGC 752.  If Hyades data are used, the data specified in Table 2 may be added.  Before that is done, however, 
users should make a point of reading the Table 2 footnotes.

A note is needed about project papers containing Hyades {\it VRI\/} data.  Users who do not require Hyades 
values of $V-R$ can limit their attention to Table 2 and CDS data files from TJJ.\footnote{Except when required for 
the sake of clarity, subscript ``C'' for Cousins indices is omitted in this paper to simplify notation.  However,
subscript ``L'' (for Landolt) is used whenever it is required.}  Those who wish to include Hyades values of $V-R$ 
should consult \citet{tj85}, \citet{jtlw06}, and \citet{jtlw08} and the footnotes given in Table 1 to the entries 
for those papers.

\subsection{Aims, procedures, and zero point uniformity}

One aim of the projects has been to determine exact relationships between their published data and the Cousins, 
Johnson, and Str{\" o}mgren-$\beta$ systems.  A second aim (as noted in \S 1) has been to establish zero points that 
are known to be uniform from cluster to cluster.  To support the latter aim, a number of nights have been used to 
compare two or three clusters directly.  These ``Sturch comparisons'' imitate a procedure applied by \citet{s72,s73}.  
In addition, zero point accuracy has been assessed from statistical comparisons of independent data sets (or ``SCIDS'' 
for short).  The mean residuals obtained from SCIDS satisfy the so-called ``FM'' (``few millimag'') standard--that 
is, they are typically a few mmag in size.  As explained in \S\S 3--6, the mean residuals show that for each magnitude 
or color of interest, zero points for the project data are coherent from cluster to cluster at the FM level.

\subsection{Proof of concept}

Both the SCIDS protocol and the FM standard have proved to be controversial.  In part, this problem is due to 
two durable axioms.  According to one of them (noted in \S 7.3 of \citealt{tj06}), the FM standard is rendered 
meaningless by a lower limit of about 10--20 mmag on the precision and accuracy of photometry.  According 
to the other axiom (noted in \S 4.4 of TJJ), there is an inescapable danger that photometric transformations 
will compromise data accuracy.  The second axiom is relevant here because SCIDS can be applied only to input 
data on a common photometric system.  For that reason, transformations must be applied to some of those data 
before they are analyzed.

To put such issues in perspective, we note first that the zero point projects are based on a strictly Baconian 
outlook (see \citealt{b1620}, Second Book, aphorism 10).  To paraphrase a relevant aphorism from that source: 
``We must not imagine or invent, but discover the properties of [the data themselves].''  That is done solely 
by applying statistical analysis to them.  In contrast, no authority is conceded to axioms like those cited
just above.

To see where the Baconian protocol leads, readers are invited to consult the first two parts of \S 11 of TJJ.  
That discussion yields some pertinent conclusions.  For one thing, no 10--20 mmag limit that would rule out
the use of the FM standard exists.  Moreover, transformations need not compromise data accuracy, even at the FM 
level.  In their Table 14, TJJ display photometric data whose zero points cohere at that level.  Additional 
examples of this sort are cited in \S\S 3.3, 3.5, and 3.7.  (For further evidence supporting the relevance of 
the FM standard, readers are invited to consult \S 7.3 of \citealt{tj06}.)

\subsection{Zero point coherence: describing results}

In the papers produced by the projects, offsets derived by applying SCIDS are quoted repeatedly.  Quite often, 
results of tests for scale factor differences are reported as well.  With one exception, however (see Table 5 
of \citealt{tj05}), no such differences are detected.  In response, only results from zero point tests are 
summarized here for the sake of brevity.

Let 

\begin{equation}
M \equiv m \pm \sigma_m
\end{equation}

\noindent refer to a formal correction to a new data set that is obtained by comparing it to an extrinsic data 
set.  In the contributing papers, values of $M$ tend to be reported in groups.  To summarize results for a given 
group, one procedure adopted here is to state a value of

\begin{equation}
t \equiv |m/\sigma_m|.
\end{equation}

\noindent This practice is based on the following lemma: if there are $N = 4$ or more values of $M$ in a given 
group, and if $t < 2.5$ for all group members, it is fair to conclude that none of them differ from zero with 
a confidence level $C \geq 0.95$.  This lemma does not depend on the value of $N$ as long as $N \geq 4$, 
and it is independent of the number of degrees of freedom for each contributing datum.\footnote{The lemma is
derived from an algorithm called false-discovery rate (or ``FDR''; see \citealt{m01}, especially \S 3 and 
Appendix B of that paper).  Unlike a simple $t = 2$ detection threshold, FDR can be applied to the results of 
either single or multiple statistical tests.  In their \S 2, \citealt{m01} discuss extensively the drawbacks of 
adopting the $t = 2$ threshold when multiple tests are performed.}

For some groups of interest, only maximum values of $t$ are reported below.  In these cases, readers should regard
Table 14 of TJJ as an example while remembering that $t < 2$ in that case.  Comments are added when a maximum 
value of $t$ exceeds 2.5 or if at least one nonzero value of $M$ is detected.  If $N < 4$ or if values of $M$ 
are scattered, informative values of $M$ are quoted instead of a limit on $t$.  For readers who desire further
information, references to the project papers are given below.  In particular, those references should be 
consulted for results of scale factor tests or to inspect all values of $M$ in particular groups.

\section{COUSINS {\it VRI\/} COLORS}

\subsection{Comparing the SAAO and 1983 Landolt systems}

Because the {\it VRI\/} projects are based on two different systems of standard stars, there is an obvious need
to understand the relationship between those systems.  This problem has been investigated by a number of authors.  
However, the meaning of their results has been obscured by use of two different conventions for reporting them.  
Authors in the southern hemisphere prefer to work with values of $V-R$ and $V-I$.  They find that for both indices, 
there are scale factor differences between the Landolt and southern hemisphere standard star data (see, for 
example, Fig.\ 2 of \citealt{mm91}).  On the other hand, authors in the northern hemisphere frequently adopt $R-I$ 
in place of $V-I$.  For $R-I$, neither a scale factor difference nor a zero point difference between the two systems
can be found (see Table 3 of \citealt{tj96}).

The choice made here is to work with $R-I$ and $V-R$.  For the latter, a standard scale factor correction is 
adopted:

\begin{equation}
(V-R)_{\rm C}=0.989(V-R)_{\rm L},
\end{equation}

\noindent with obvious notation (see \S 4.4 of \citealt{mm91} and eq.\ [5] of \citealt{tj96}).  If possible, mean 
residuals in $V-R$ and $R-I$ are reported.  When only values of $V-I$ are to be tested, equation (3) is applied 
(if necessary) and comparison data are then calculated by adding together values of $V-R$ and $R-I$.

\subsection{Instrumental systems}

The first values of Cousins $V-R$ and $R-I$ produced by this project were published by \citet{tj85}.  Those
authors combined results from a variety of instrumental systems, with some requiring special transformation 
relations (see their \S II).  Though this procedure was successful (see especially \S\S 3.4 and 3.5), it
was also a stopgap device.  Subsequently, a limited number of instrumental systems have been used instead.  
The scale factors of those systems do not differ from those of the adopted systems of standard stars by more
than 10 percent.  For the SAAO system, there are no differences as large as 5 percent.  (For detailed 
information about the scale factors, see Table 1 of \citealt{jtlw06}.)

\subsection{Hyades $R-I$}

For the Hyades (and the other program clusters as well), values of $R-I$ have been used to calculate 
temperatures.  Moreover, the accuracy of some of the Hyades data has been challenged (see \S 3.8).  For these
reasons, results from each step used to test the data will be reported in some detail.

In \citet{tj85}, eight data sets are combined.  For six of them, results of tests for zero point coherence 
are reported.  Here, it is found that $t < 1.2$ (see Tables I and IV of \citealt{tj85}).

\citet{tj05} perform data tests in two steps.  They find that measurements made by \citet{m67} differ from 
those of \citet{tj85}, with $|M| = +11.8 \pm 2.9$ mmag.  That difference affects a $V-I$ catalog published by 
\citet{pths04} because its entries are based on transformations of published photometry, with the Mendoza
data included.  If those data and the Pinsonneault et al.\ catalog results are set aside, tests using 
measurements from four extrinsic sources yield zero point coherence with $t < 1.8$.\footnote{We note that 
\citet{clhw11} have adopted corrected entries from the Pinsonneault et al.\ catalog.  Because data in that 
catalog have low precision, this is not the best available procedure.  For a detailed critique of the 
Pinsonneault et al. catalog, see \S\S 1, 2, and 4.3 of \citealt{tj05}.  For Hyades data with acceptable 
precision, see the Hyades catalogs of TJJ.}  (For further information about those tests, see Table 1 of 
\citet{tj05}.)

In their second step, \citet{tj05} compile an $R-I$ catalog using data from the 1985 paper and other sources 
(listed in Table 4 of \citealt{tj05}).  Here also, an offset is found for the Mendoza data, as one might 
expect.  However, data sets from five other extrinsic sources yield zero point coherence with $t < 2.0$.
Except for the Mendoza data, there is no overlap between the extrinsic data sources used in this step and 
those used in the tests applied in the first step.  (Sources of catalog data and pertinent values of $M$ are 
listed in Tables 4 and 5 of \citealt{tj05}, respectively.  Because a revised normalization is used for the 
Mendoza data, the result is a larger derived offset for those data than before: $|M| = +36.0 \pm 5.7$ mmag.)

Formal corrections to the 1985 and 2005 data are reported in Table 3 of \citet{jtlw06}.  Those corrections
have been derived using SAAO data.  For the epoch 1985 measurements, the SAAO data yield $M = -0.9 \pm 1.0$ 
mmag.  For the epoch 2005 catalog data, $M = +0.1 \pm 1.0$ mmag.  Given the latter result, the catalog data 
and the SAAO data can be combined readily.  The resulting means are described in \S 10.1 of TJJ.

\subsection{Hyades $V-R$}

In this case, as for $R-I$, there is a published accuracy challenge (see \S 3.8).  In response, two steps 
are taken.  \citet{tj85} list seven contributing data sets, and they report the results of zero point
consistency tests for all seven of them (see their Tables I and IV, respectively).  Here, it is found that 
$t < 2.3$.

The results of $V-R$ tests performed by \citet{tj05} are not reviewed here because they are ambiguous (see
\S 4.2 of \citealt{tj05}).  In \citet{jtlw06}, both SAAO data and measurements from the 2005 paper are brought 
to bear on the zero point problem.  For data from \citet{tj85}, the derived formal correction $M$ is
$-3.5 \pm 1.3$ mmag.  Though $t = 2.7$ for this residual, an analysis using false discovery rate reveals 
that it does not differ from zero with $C \geq 0.95$.

\subsection{M67}

TJJ present M67 data that are largely derived from new measurements.  For a published source that contributes 
to the catalog, there is an accuracy challenge (see \S 3.8).  Here, zero point comparisons are made between the 
catalog data and 13 other data sets, with $N = 6$, 4, and 3 for $R-I$, $V-R$, and $V-I$, respectively.  If $V-I$ 
measurements by \citet{s04} and \citet{mmj93} are set aside, it is found that $t < 2.5$ for the other 11 tests, 
with those using SAAO data being included.  Judging from this consensus, the zero points of the TJJ data appear
to be reasonably secure.

The Sandquist $V-I$ data yield a formal correction $M$ to the TJJ data of $-4 \pm 1$ mmag.\footnote{Here and for
the Montgomery et al.\ data, quoted values of $M$ differ from those in Table 5 of TJJ because the latter do not 
include an adjustment of 2 mmag.  That adjustment is described in \S 5.3 of TJJ.}  This offset does not allow
one to be sure that the zero point of the Sandquist data differ from those of the E region standards (see \S 6
of TJJ).  However, the existence of the offset should not be overlooked.

For the Montgomery et al.\ values of $V-I$, evidence is found for a scale factor offset.  In addition, the 
Montgomery et al.\ results are more positive than those of TJJ by $27 \pm 1.3$ mmag.  Allowance for that offset 
appears to resolve a puzzling differential zero point problem discussed by \citet{vdbc03} and \citet{vdbs04} 
(see \S 7 of TJJ). All told, there is a good case for not using the Montgomery et al.\ data in the future.  
(For more about the Montgomery et al.\ data, see \S\S 4.3 and 5 of this paper.  For details about the M67 
data testing, see \S\S 5.3 and 6 of TJJ and \S 5 of \citealt{jtlw08}.)

\subsection{Coma}

This cluster is too far north to be observed at SAAO.  Partly for that reason, only one $V-R$ result from
SCIDS is available.  Using data from \citet{tj05}, the value of $M$ for the \citet{tj85} data is found to 
be $+2 \pm 1.5$ mmag.

For $R-I$, the Coma data of \citet{tj85} are based on a Sturch comparison of Coma, M67, and the Hyades (see 
\citealt{t78}).  Judging from an analysis of the resulting M67 data, the maximum correction that could be 
required for the Coma data of \citet{tj85} is about 2 mmag (see \S 5.3 of TJJ).  In fact, two tests of the 
Coma data yield $M = -1 \pm 1.5$ mmag and $M = +2 \pm 1.4$ mmag, respectively.  (For further information, 
see \S 8 of TJJ.)

\subsection{Praesepe and NGC 752}

For these clusters, there are not as many data sets as there are for the Hyades or M67, so not as many
tests can be performed.  In addition, only one of the clusters (namely, Praesepe) can be observed from 
SAAO.  However, since Sturch comparisons of the clusters were performed frequently, SAAO data can be used 
to test the measurements for both of them.  For $V-R$, two comparisons can be made, yielding $M = +3 \pm 
2.3$ mmag and $ M = -2.9 \pm 1.3$ mmag for the two clusters (see Table 7 of TJJ and Table 2 of 
\citealt{jtlw11}).

For values of $R-I$ for dwarfs, there is one discrepant value of $M$ ($+6 \pm 1.4$ mmag).  However, when SAAO 
data are included, $t$ is found to be $< 1.7$ for four other such values.  For $R-I$ measurements for giants 
in NGC 752, derived values of $M$ are $+9 \pm 2.7$ mmag and $-5 \pm 2.9$ mmag.  The latter result is 
thought to be the more trustworthy of the two because it relates a formal zero point correction for NGC 752 
to the well-supported formal correction for M67 (see \S 3.4).  The $R-I$ results for NGC 752 are therefore 
accepted, though with an acknowledgment that they should eventually be tested further (see Table 2 of 
\citealt{jtlw11} and \S 8 and Table 7 of TJJ).\footnote{Because of an editing oversight, there are references 
in \S 8 of TJJ to two ``boldface'' entries in Table 7 that are not actually given in boldface.  In the order 
in which they are encountered in the discussion of TJJ, the entries in question are the third from the last 
and second from the last entries in the table.}

\subsection{Accuracy challenges}

To our knowledge, there are three extant challenges to the accuracy of our {\it VRI\/} colors.  One of them 
is put forward in \S 2.2 of \citet{vdbs04}.  The salient objectionable features of that challenge are $a)$ 
use of data that have been described in print, but not published, $b)$ use of data whose accuracy is 
irrelevant to our measurements, and $c)$ substitution of inspection of graphs for statistical analysis.  
Responses to that challenge appear in \S\S 4.1 and 12 of TJJ.

A second challenge concerns the \citet{tj85} transformations from instrumental to standard systems.  As 
\citet{jtlw06} note in their \S 1, the accuracy of those transformations has been criticized.  There is a key 
underlying assumption: if unsupported, qualitative objections are made to data provenance, the data themselves 
have to be inaccurate.  Here, the axial issue is what the data themselves say.  Alleged problems with the 
Taylor \& Joner transformations cannot explain the coherence of their output--to say nothing of the support 
offered by subsequent measurements (see \S\S 3.3 and 3.4).  

A third challenge, offered by \citet{a07}, has not previously been assessed in the literature.  Those authors 
assume axiomatically that only data obtained with a single instrumental system can be trustworthy.  On that 
basis, An et al.\ (see their Appendix) adopt SAAO Hyades photometry from \citet{jtlw06} while rejecting 
the multi-source catalog of \citet{tj05}.  Here also, the axial issue is what the data themselves say.  An et 
al.\ do not acknowledge that the data they accept have been used in successful SCIDS tests of the data they 
reject (again see \S\S 3.3 and 3.4).  Joner et al.\ highlight the results of those tests (see their \S 4.2) and 
list them in their Table 3.  However, the only table in that paper considered by An et al.\ is Table 2, in which 
the Joner et al.\ photometric data are listed.  Speaking as two of the co-authors of \citet{jtlw06}, we find 
such selective citation of that paper to be open to serious question.

In assessing these criticisms, we have given priority to the published data themselves.  In contrast, those 
who employ provenance reasoning or axioms have conferred ultimate priority on them.  Judged in the context of 
the Baconian standard (recall \S 2.3), the resulting criticisms are not convincing.  Nevertheless, it must be 
acknowledged that within the photometric discipline, the Baconian standard itself is apparently controversial.

Another concern has been our applications of statistical analysis to photometric data.  In our correspondence 
with other photometrists, this issue has provoked the most forceful criticisms of the zero point projects that 
we have encountered.  The instance put forward by \citet{vdbs04} is actually a mild example of rejection of 
statistical analyses.  As a rule, such rejections have taken place if the results of the analyses contradict 
axioms that are deemed to have priority.  These rejections help to underscore the controversial character
of the Baconian standard.  (The outlook that would follow from general adoption of that standard is discussed 
in \S 7.)

\subsection{Adopted sources of reddening values}

In some CDS files produced by TJJ and \citet{jtlw11}, data corrected for reddening (if necessary) are given.  
For that reason, reddening values must be known for all the program clusters.  The reddening values 
adopted for the Hyades, Coma, and Praesepe have been derived by \citet{t06}.  The adopted values for M67 
and NGC 752 are given by \citet{t07a} and \citet{t07b}, respectively.

For Praesepe and NGC 752, problems were encountered in deriving final reddening values.  To assess the first of
those problems, users of Praesepe data are advised to consult \S 9.1 of \citet{t06}.  A quick way to assess the 
problem for NGC 752 is to look at the first two entries in Table 2 of \citet{t07b}.  A full discussion of that
problem is given in \S 5 of that paper.

The M67 reddening value adopted here is not accepted universally or without doubt (see, for example, 
\citealt{y08}).  For that reason, readers are invited to gauge its rationale by consulting Tables 1 through 
3 of \citet{t07a}.  As those tables show, almost all M67 reddening work done before 2007 did no more than
produce 70 values of $E(B-V)$ that range over 0.14 mag.  Some authors still adopt one or more of those 
results without explanation (see \citealt{y08,f10}).  That practice is misleading because it effectively 
conceals the range of the 70 reddening values in the parent population.  There is no reason to suppose that 
unexplained selections from those data bear any genuine relation to the actual M67 reddening value.  By 
consulting Table 6 of \citet{t07a}, readers can see what was accomplished by adopting a fresh approach to 
the reddening problem.  It is suggested that after inspecting Tables 1 through 3, users of M67 photometry 
disregard their contents altogether and adopt $E(B-V) = 41 \pm 4$ mmag from Table 6.

\section{$V$ MAGNITUDES FROM {\it BVRI\/} PHOTOMETRY}

\subsection{Zero point comparisons: a first inspection}

To assess $V$ magnitude consistency, it is useful to start with Tables 8 and 9 of TJJ.  The values of $M$ 
given there range over about 60 mmag.  The largest value of $M$ listed by TJJ is based on data from \citet{jk55}.
The zero points for the measurements of Johnson \& Knuckles are not definitive (for an explanation, see \S
7 of \citealt{jtlw08}).  However, if values of $M$ derived for their data are set aside, the range of the 
remaining values of $M$ drops only to about 40 mmag.  Here, then, no examples of consensus like those found 
for {\it VRI\/} colors appear.  The task at hand is to satisfy the FM standard despite this obstacle.

\subsection{Results for epoch 2011}

The next step is to consult \citet{jtlw11}.  One issue discussed by those authors is the relationship between
$V$ magnitudes on the Johnson system and $V$ magnitudes on the SAAO and \citet{l83} systems.  No color 
correction of SAAO $V$ magnitudes appears to be required, and best evidence suggests that this is also true 
for \citet{l83} $V$ magnitudes (see \S 4 and Table 3 of \citealt{jtlw11}).  The derived formal zero point 
correction for the latter (called $M_3$ here) is $+4.4 \pm 2.4$ mmag.  Since the corresponding value of $t$ 
is $< 2$, $M_3$ does not differ from zero with $C \geq 0.95$.  The corresponding result for SAAO $V$ 
magnitudes is derived in two ways, with the results being $+6.8 \pm 1.1$ and $+2.0 \pm 0.8$ mmag, respectively.  
Though the first of these values is preferred, it is not regarded as definitive (see \S 5 and Table 4 of 
\citealt{jtlw11}).

\citet{jtlw11} also consider Hyades, M67, and Praesepe $V$ magnitudes measured in the northern hemisphere.  
It is found that in effect, those data are on a single zero point (see \citealt{jtlw11}, Table 4, entries 
3--5).  This deduction should carry some weight because it seems unlikely to be an artifact of coincidence.
Encouragingly, the formal difference between the three-cluster zero point and that of the Johnson system is 
the null value of $M_3$ quoted above.

\subsection{Interpreting collected results}

The final step is to interpret values of $M$ collected from TJJ and \citet{jtlw11}.  For Coma, tests of data in
the `Landolt'' group yield $M = -6 \pm 4$ mmag and $M = +2 \pm 3$ mmag.  The latter value actually applies for 
combined Hyades and Coma measurements, and it is encouragingly consistent with the value of $M_3$ quoted just
above.  (The two quoted values of $M$ are from the fourth and fifth entries in Table 8 of TJJ.)

For Praesepe and NGC 752, results may be treated as a unit (as noted in \S 3.7).  Here, TJJ could not
come to a definite conclusion because their derived values of $M$ are scattered.  With $M_3$ added, however, 
the balance of evidence favors a null $V$ correction for both clusters.  (For the results available to TJJ, 
see the last six entries in their Table 8.)

For M67, TJJ list eight values of $M$ in their Table 9, with the largest being $+24 \pm 1$ mmag.  However, only 
the last entry in that table relates the TJJ $V$ data directly to the Johnson system.\footnote{In the second from
last sentence in their \S 9, TJJ refer to that entry as a boldface entry.  In fact, it is not in boldface because
of a proofreading error.  However, the TJJ reference does designate the correct entry in their Table 9.}  The 
implied correction to the TJJ data is formally zero, but has a $2\sigma$ uncertainty of 12 mmag.  With $M_3$ 
added, the implied correction is still formally zero, but the $2\sigma$ uncertainty drops to 4.8 mmag.

It should be noted (see Table 1 of TJJ) that an error that correlates with position on the face of the cluster
has been detected in $V$ magnitudes from \citet{mmj93}.  There are also problems with values of $V-I$ and 
$B-V$ from that source (see \S\S 3.5 and 5, respectively).  All told, there is good reason to refrain 
altogether from using the Montgomery et al.\ {\it BVRI\/} data in the future.

\section{$B-V$ COLORS AND VALUES OF [Fe/H]}

Color-magnitude analyses may include values of $B-V$ as well as {\it VRI\/} photometry.  For that reason, it is 
worthwhile to specify $B-V$ sources for the program clusters.  As it happens, the ``L83'' papers (see Table 1) 
include no $B-V$ photometry.  Fortunately, there are useful results in the ``SAAO'' and other papers.

Typically, there are only two or three $B-V$ data sources of interest for each of the program clusters.  An 
exception is M67, for which there are seven sources with values of $|M|$ ranging up to about $+25$ mmag.  A 
zero point for those data has been derived from SAAO measurements and results from a Sturch comparison (see 
\citealt{s73}).  Users who are interested in further information about the zero point procedure applied in 
this case are invited to consult \S 8 of \citet{jtlw08}.

The $B-V$ data sources we recommend for use are listed in Table 3.  Instead of the results of comparisons among
those sources, the table lists formal zero point corrections for each of them.  Users who are interested
in the provenance of those corrections are encouraged to consult either the footnotes to Table 3 or sources
cited in the body of the table.  If necessary, reddening corrections from sources listed in \S 3.9 should be
applied to the adopted values of $B-V$.  

Two data sources of potential interest are not listed in Table 3.  \citet{m67} data are excluded because more
precise results are available from sources listed in Table 3.  $B-V$ measurements given by \citet{mmj93} are 
excluded because they appear to be on two decisively different zero points (see \S 8 and entries 6 and 7 in 
Table 5 of \citealt{jtlw08}).  {\it VRI\/} results from Montgomery et al.\ are also not recommended for future 
use (see \S\S 3.5 and 4.3).

Table 3 includes two entries for Hyades data sets.  However, we recommend that for single Hyades stars, the 
following convenient equations be used instead (see \S 6 of \citealt{jtlw08}):

\begin{equation}
B-V = \sum_{i=0}^3 C_i r^i,
\end{equation}

\noindent with $r \equiv (R-I)_{\rm C}$; 

\begin{equation}
[C_0, C_1] = [(0.244 \pm 0.001), (-1.13 \pm 0.41)];
\end{equation}

\noindent and

\begin{equation}
[C_2, C_3] = [(9.53 \pm 1.45), (-7.89 \pm 1.61)],
\end{equation}

\noindent with $0.11 \mkern6mu {\rm mag} \leq (R-I)_{\rm C} \leq 0.50 \mkern6mu {\rm mag}$.

Because $B-V$ measurements are sensitive to blanketing, they must be analyzed with the help of cluster metallicities.  
For four of the program clusters, values of [Fe/H] that were current at time of publication are given in the reddening 
papers cited in \S 3.8.  For Praesepe, an updated value of [Fe/H] appears in \citet{t08}.  It should be noted that 
these values of [Fe/H] apply for temperatures on an angular diameter scale (see \citealt{t07a}, Appendix B, entry 4).  
Though the corrections to published data that are required to put them on that scale are often small, it should not 
be forgotten that those corrections have been applied.  For that reason, the resulting values of [Fe/H] are not 
strictly comparable to counterparts in the literature.

\section{THE STR{\" O}MGREN-$\beta$ PROJECT}

When this project began, it was known that there are appreciable zero point differences between a number of published 
Str{\" o}mgren-$\beta$ data sets (see, for example, the Appendix of \citealt{ntc87}).  To deal with such problems,
the project aims to ``$. \mkern6mu . \mkern6mu .$ test the consistency of Str{\" o}mgren-$\beta$ photometry for 
lightly reddened galactic clusters'' (see \S 1 of \citealt{jt97}).  To make that possible, Sturch comparisons have
been used to build up a system of standard-star data on a uniform zero point.  That system is based on measurements of 
the Hyades and Coma, for which extant data (\citealt{cp66,cb69}) turn out to share a common zero point.  The Hyades-Coma 
paper (see \citealt{tj92}) and three others produced by the project are listed in Table 1.  \citet{jt07} have presented 
comparable results for the Pleiades, and \citet{jtpj95} have added an auxiliary paper on the photometry of 
\citet{go76,go77}.

Relative to the Hyades-Coma system, offsets have been detected for M67, Praesepe, and NGC 752.  For the latter two 
clusters, all but one of the derived values of $|M|$ (with $M$ being a mean residual; see eq.\ [1]) range from 
zero to $+19.0 \pm 3.4$ mmag.  For M67, values of $|M|$ as large as $+61.0 \pm 5.8$ mmag have been found (see Tables 
4 and 5 of \citealt{jt97} and Table 6 of \citealt{jt07}.)  To date, data from the project have been used in zero point 
tests and color-color diagnosis (see, for example, \citealt{att06} and \citealt{k98}, respectively).  Use of project 
results to establish cluster measurements as standard-star data is also possible (see, for example, Appendix B of 
\citealt{jt95}).

\section{PERSPECTIVES}

To grasp the overall results of the zero point projects, suppose first that the Baconian standard (recall \S 2.3) 
were to prevail throughout the photometric discipline.  As one outcome, statistical analysis would be applied
dependably to photometric data sets, and the results would take precedence over all contrary axioms about the 
alleged character of the data (see \S 3.8).  A second outcome would be increased caution about gauging photometric 
quality.  Too often, discourse among photometric astronomers takes place in an eye-smarting haze of skepticism 
about the quality of published photometry.  Inventing poorly supported and often specious criticisms of such data 
is a widespread practice (see, for example, objections discussed in \S 3.8 and a class of criticisms diagnosed in 
\S 7 of \citealt{tj05}).  

After the changeover to the Baconian standard, a lesson taught by photometric zero points would attract notice.  It 
would be acknowledged that maximum values of $|M|$ can be about 86 mmag (see Table 3 of \citealt{smv04}), 40--60 mmag 
(see \S\S 4.1 and 6), 25 mmag (see section 5), and $\leq 0.9 \pm 1.0$ mmag (see \S 3.3).  In response, there would be 
no generalizations about any alleged limit to zero point coherence (see notably \citealt{sbg03} and \S 3 of 
\citealt{smv04}).  Instead, particular data sets would be gauged solely on their particular merits and demerits.

Yet another result would be a reluctance to ``block the road of inquiry'' (to quote Charles Sanders Peirce).  It 
would be recognized that skepticism about published data is not conducive to putting them to good use.  Ways of
doing that would receive greater attention and would be more widely practiced.  Using extant photometry to derive and 
apply standard corrections to data sets (see \S\S 3.1 and 5) is one example of such practice.  The \citet{m67} data 
offer another example: if Hyades measurements are used to calculate an auxiliary transformation for those data, 
measurements of other clusters made by Mendoza can then be compared usefully to counterparts from other sources 
despite the relatively low precision of the Mendoza data.  (For the auxiliary transformation, see eqs.\ [A5] in 
Table 9 of \citealt{tj05}.  For a fully successful use of the Mendoza data, see Table 7 of TJJ.)  Finally, zero
point jitter in existing data for given clusters would not be accepted indefinitely.  Instead, Sturch comparisons 
would be employed, as in an example to be given just below.

To potential users, we suggest three ways of using data from the zero point projects.  There is sizeable zero 
point jitter in extant photometry for NGC 188 (see again Table 3 of \citealt{smv04}).  To derive reliable zero points 
for that photometry, Sturch comparisons of NGC 188 and M67 could be performed at equal air mass.  The results could 
then be reduced by using M67 data from TJJ.  In addition, a test could be performed to see whether $V-I$ measurements 
for the Hyades and M67 can be matched simultaneously to theoretical isochrones (see \S 3.5).  Finally, statistical 
algorithms for color-magnitude analysis (see, for example, \citealt{jl05}) could be applied to project data.  The aim 
would be to derive ages and distance moduli for the program clusters with the highest possible precision.

\acknowledgments

With alacrity, we thank our collaborators--Dr.\ Elizabeth J.\ Jeffery, Dr.\ C.\ David Laney, and Francois van Wyk--for 
their invaluable work on the zero point projects.  We add that judgments offered in this paper are ours specifically
and are not necessarily shared by our collaborators.

\clearpage


\begin{deluxetable}{lccllccl}
\tabletypesize{\scriptsize}
\tablenum{1}
\tablewidth{0pt}
\tablecaption{List of project {\it VRI\/} and Str{\" o}mgren-${\beta}$ papers}
\tablehead{
 & & \colhead{Use} & & & & \colhead{Use} & \\
Paper & \colhead{Group\tablenotemark{a}} & \colhead{data?} & \colhead{Clusters\tablenotemark{b}} & 
Paper & \colhead{Group\tablenotemark{a}} & \colhead{data?} & \colhead{Clusters\tablenotemark{b}} 
}
\startdata
Taylor (1978)\tablenotemark{c}          & L83  & N & H,C,M67 & TJJ   & L/S  & Y\tablenotemark{d}     & 5 \\
Taylor \& Joner (1985)                  & L83  & Y\tablenotemark{e} & H,C,M67 & Joner et al.\ (2008) & SAAO & Y\tablenotemark{f} & H,M67  \\
Joner \& Taylor (1988)\tablenotemark{g} & L83  & N & H,M67   & Joner et al.\ (2011)   & SAAO & Y\tablenotemark{h} & H,Pr   \\
Taylor \& Joner (1988)                  & L83  & N & H,M67   & Taylor \& Joner (1992) & Str  & Y\tablenotemark{i} & H,C    \\
Joner \& Taylor (1990)\tablenotemark{g} & L83  & N & M67     & Joner \& Taylor (1995) & Str  & Y\tablenotemark{i} & Pr,752 \\
Taylor \& Joner (2005)                  & L83  & N & H       & Joner \& Taylor (1997) & Str  & Y\tablenotemark{i} & M67    \\
Joner et al.\ (2006)                    & SAAO & Y\tablenotemark{j} & H & Joner \& Taylor (2007) & Str & N\tablenotemark{k} & Pr \\
\enddata
\tablenotetext{a}{``L83'': Cousins {\it VRI} reduced to the Landolt (1983) standard-star system.  ``SAAO'': 
$B-V$ and Cousins {\it VRI}, South African Astronomical Observatory.  ``L/S'': combined L83 and SAAO measurements.
``Str'': Str{\" o}mgren-$\beta$ measurements.}
\tablenotetext{b}{''H,'' ''C,'' ''M67,'' ''Pr,'' and ''752'' refer to the Hyades, Coma, M67, Praesepe, and NGC 752,
respectively.  ``5'' means that all five clusters are considered.}
\tablenotetext{c}{To see how data from this paper have been used, look at the entry for run 3 in \S II of Taylor \&
Joner (1985).}
\tablenotetext{d}{For the Hyades, CDS files for this paper contain combined northern-hemisphere and SAAO data.  For all 
program clusters except Praesepe, CDS files from TJJ should be consulted.  For Praesepe data, see Joner et al.\ (2011).}
\tablenotetext{e}{Despite the headings in Table V of Taylor \& Joner (1985), values of $(V-R)_{\rm L}$--not 
$(V-R)_{\rm C}$--are given there.  Hyades values of $(V-R)_{\rm L}$ data from Table V may be adopted if they are not 
flagged with an asterisk.  They may then be converted to Cousins values of $V-R$ by applying eq.\ [3].}
\tablenotetext{f}{The TJJ M67 measurements are adequately precise without being combined with data given by Joner et al.\ 
(2008).  For Hyades data from that paper that are to be added to the TJJ CDS data files, see Table 2.  Hyades values 
of $V-R$ are available from Joner et al.\ (2008).}
\tablenotetext{g}{Both the data and the discussion in the listed paper have been superseded.}
\tablenotetext{h}{A CDS file is available containing combined Praesepe data from TJJ and Joner et al.\ (2011).  The TJJ
Hyades measurements are adequately precise without being combined with data given by Joner et al.\ (2011).}
\tablenotetext{i}{The listed paper includes newly published Str{\" o}mgren-$\beta$ measurements.  These may be combined
with photometry (corrected, if necessary) from previously published sources cited in the listed paper.}
\tablenotetext{j}{Use only Hyades $V-R$ measurements from Joner et al.\ (2006).  For other Hyades data, see TJJ and Table
2 of this paper.}
\tablenotetext{k}{Joner \& Taylor (2007) do not give new data for the five clusters of interest here, but do give such 
data for the Pleiades.}

\end{deluxetable}

\clearpage


\begin{deluxetable}{cccccc}
\tabletypesize{\scriptsize}
\tablenum{2}
\tablewidth{0pt}
\tablecaption{Additions to the TJJ Hyades data}
\tablehead{
\colhead{vB\tablenotemark{a}} & \colhead{HIP} & \colhead{$R-I\tablenotemark{b}$} & 
\colhead{$(V-K)_{\rm J}\tablenotemark{c}$} & \colhead{$\theta\tablenotemark{d}$} & \colhead{$V\tablenotemark{e}$} 
}
\startdata
231 & 19207 & 0.583(3.0) & 2.764(11) & 1.147(2.1) & 10.469(3.0) \\
253\tablenotemark{f} & $-$ & $-$ & $-$ & $-$ & $-$ \\
262 & 20527 & 0.665(3.0) & 3.067(11) & 1.201(1.9)\tablenotemark{g} & 10.876(4.5) \\
291 & 21261 & 0.596(3.0) & 2.812(11) & 1.155(2.0) & 10.691(5.2) \\
311 & 21723 & 0.516(4.2) & 2.516(16) & 1.098(3.1) & 10.024(10.) \\
324 & $-$   & 0.508(3.0) & 2.486(11) & 1.092(2.2) & $\phantom{1}9.827(4.5)$ \\
$-$ & 19808 & 0.637(3.0) & 2.964(11) & 1.183(1.9) & 10.679(2.7) \\
\enddata
\tablenotetext{\ }{NOTE.--These data are not in the TJJ CDS data file for the Hyades.  Before deciding to add
them to that file, users are urged to be sure that the character of the data is understood.}
\tablenotetext{a}{This is the van Beuren (1952) number.}
\tablenotetext{b}{This entry is from Table 1 of Joner et al.\ (2008).  The entries in parentheses are standard
errors in mmag with an adopted lower limit of 2.9 mmag.}
\tablenotetext{c}{The entries in parentheses are standard errors in mmag.  For the sake of consistency with
existing results, all quoted values of $(V-K)_{\rm J}$ have been transformed from the values of $R-I$ in the 
third column using a relation given in Table 4 of TJ06.  Adopting the modification for red stars given in the 
Appendix of Joner et al.\ (2008) would change the resulting values of $(V-K)_{\rm J}$ by less than $1\sigma$.}
\tablenotetext{d}{As usual, $\theta \equiv 5040/T_{eff}$. The entries in parentheses are standard deviations
multiplied by 1000.}
\tablenotetext{e}{These entries are from Table 1 of Joner et al.\ (2008), with a compromise correction of 4
mmag added (see Joner et al.\ (2011), Table 4, entries 11 and 13).}
\tablenotetext{f}{There is a $3\sigma$ difference between the values of $R-I$ given for this star by TJJ and 
Joner et al.\ (2008).  The TJJ entry for this possible variable star should therefore be deleted.}
\tablenotetext{g}{This value of $\theta$ is from an extrapolation of the Di Benedetto (1998) temperature scale
and so should be used with caution.}
\end{deluxetable}


\begin{deluxetable}{llcccccc}
\tabletypesize{\scriptsize}
\tablenum{3}
\tablewidth{0pt}
\tablecaption{Recommended $B-V$ sources for program clusters}
\tablehead{
 & & \colhead{Correction} & & \colhead{Source} & & \colhead{Error} \\
Cluster & Paper\tablenotemark{a} & \colhead{(mmag)} & \colhead{Apply?} & \colhead{paper\tablenotemark{b}} & \colhead{Table\tablenotemark{c}} 
& \colhead{(mmag)} & \colhead{Kind\tablenotemark{d}}
}
\startdata
Hyades   & JK55     & $-3.5 \pm 1.3$\tablenotemark{e} & Y & SAAO II  & Table 4 & 10\tablenotemark{f} & R \\
Hyades   & SAAO I   & $-9.4 \pm 1.5$\tablenotemark{e} & Y & SAAO II  & Table 4 & $-$ & I \\
M67      & SAAO II\tablenotemark{g} & $+2.3 \pm 2.4$  & N & SAAO II  & Table 5 & $-$ & I \\
M67      & S04      & $+2.3 \pm 2.4$                  & N & SAAO II  & Table 5 & 13  & S \\
Praesepe & J52,DKK\tablenotemark{h} & $-2.9 \pm 3.0$  & N & SAAO III & Table 2 & 10\tablenotemark{f} & R \\
Praesepe & SAAO III & $-9.4 \pm 1.5$                  & Y & SAAO II  & Table 4 & $-$ & I \\
Coma     & JK55     & $-2.3 \pm 2.7$                  & N & New\tablenotemark{i} & $-$ & 10\tablenotemark{f} & R \\
NGC 752  & J53      & $-14.3 \pm 3.1\phantom{1}$      & Y & New\tablenotemark{i} & $-$ & 10\tablenotemark{f} & R \\
NGC 752  & E63      & $-9.4 \pm 3.2$                  & Y & New\tablenotemark{i} & $-$ & 10  & R \\
\enddata
\tablenotetext{a}{These papers present the data to be adopted.  ``DKK'' is Dickens et al.\ (1968); ``E63'' 
is Eggen (1963); ``J52'' is Johnson (1952); ``J53'' is Johnson (1953); ``JK55'' is Johnson \& Knuckles 
(1955); ``S04'' is Sandquist (2004); ``SAAO I'' is Joner et al.\ (2006); ``SAAO II'' is Joner et al.\ 
(2008); and ``SAAO III'' is Joner et al.\ (2011).}
\tablenotetext{b}{These are the source papers for the quoted corrections.  ''SAAO II'' is Joner et al.\ 
(2006); ``SAAO III'' is Joner et al.\ (2011); ``New'' refers to this paper.}
\tablenotetext{c}{Each listed table appears in the source paper in the previous column.}
\tablenotetext{d}{``I'': standard errors are quoted in the data table. ``R'': the quoted error is an rms
error.  ``S'': the quoted error is a standard error.}
\tablenotetext{e}{For a preferred source of $B-V$, see eqs.\ [3]--[5] in the text.}
\tablenotetext{f}{Johnson implies that the rms error for his data is 7 mmag.}
\tablenotetext{g}{See Table 6 of the listed paper.  Because rezeroed SAAO measurements from Table 2 of 
that paper contribute to its Table 6, no entry for the Table 2 data is given here.}
\tablenotetext{h}{Data from the two listed papers are effectively on the same zero point; see Joner et al.\ 
(2011), Table 2, entry 6.}
\tablenotetext{i}{This correction is derived by using extrinsic data from eq.\ [1] of Joner et al.\ (2006).  For
source data, see measurements and sources quoted by Taylor \& Joner (1992) and Joner \& Taylor (1995).}

\end{deluxetable}

\end{document}